\documentclass[prc,aps,twocolumn,showpacs,superscriptaddress,preprintnumbers]{revtex4-1}
\usepackage{graphicx}
\usepackage{amsmath}
\usepackage{ulem} 
\usepackage[dvipsnames]{xcolor}

\usepackage{enumitem} 
\usepackage{hyperref}
\hypersetup{backref,colorlinks=true,urlcolor=blue,linkcolor=blue,citecolor=blue}
\newcommand{\be}{\begin{eqnarray}}
\newcommand{\ee}{\end{eqnarray}}
\newcommand{\non}{\nonumber \\}

\begin{document}

\title{Correlations of $\boldsymbol{\pi N}$ Partial Waves for Multi-Reaction
Analyses}

\author{M.\ D\"oring}
\email{doring@gwu.edu}
\affiliation{
Institute for Nuclear Studies; Astronomy, Physics, and Statistics Institute of Sciences; 
Department of Physics,
The George Washington University,
725 21$^\text{st}$ St, NW, Washington, DC 20052, USA}
\affiliation{Thomas Jefferson National Accelerator Facility, 12000 Jefferson
Ave, Newport News, VA 23606, USA}
\author{J.\ Revier}
\affiliation{
Department of Physics,
The George Washington University,
725 21$^\text{st}$ St, NW, Washington, DC 20052, USA}
\author{D.\ R\"onchen}
\affiliation{
Helmholtz-Institut f\"ur Strahlen- und Kernphysik (Theorie) and\\ Bethe
Center for Theoretical Physics, Universit\"at Bonn, 53115 Bonn, Germany
}
\author{R.\ L.\ Workman}
\email{rworkman@gwu.edu}
\homepage[\\SAID web page: ]{http://gwdac.phys.gwu.edu/}
\affiliation{
Institute for Nuclear Studies and
Department of Physics,
The George Washington University,
20101 Academic Way,
Ashburn, VA 20147, USA}


\preprint{JLAB-THY-16-2320}

\begin{abstract}
In the search for missing baryonic resonances, many analyses include data from 
a variety of pion- and photon-induced reactions. For elastic $\pi N$
scattering,  however, usually the partial waves of the SAID or other groups are
fitted,  instead of data. We provide the partial-wave covariance matrices needed
to perform  correlated $\chi^2$ fits, in which the obtained $\chi^2$ equals the
actual $\chi^2$  up to non-linear and normalization corrections. For any analysis
relying on partial waves  extracted from elastic pion scattering, this is a
prerequisite to assess the  significance of resonance signals and to assign any
uncertainty on results. The influence of systematic errors is also considered.
\end{abstract}

\pacs{11.80.Et, 11.80.Gw,13.75.Gx,13.85.Dz }

\maketitle

\section{Introduction and Motivation}

The existence and properties~\cite{PDG}  of most $N$ and $\Delta$ resonances
have been determined through  elaborate  analyses~\cite{WI08, fit06, fit95,
coulomb, hoehler, cutkosky1, cutkosky2, cutkosky3} of $\pi N$ elastic scattering
data. More recently, however, baryon spectroscopy has been driven by the
progress made in the measurement and analysis of meson photoproduction
reactions. These analyses often take a  multi-channel approach, incorporating
reactions with a variety of initial ($\pi N$, $\gamma N$) and final ($\pi N$,
$\eta N$, $K\Lambda$, $K\Sigma$, $\omega N$, $\pi\pi N$) states.

In order to build on the progress made in the earlier $\pi N$ elastic analyses,
multi-channel analyses~\cite{Ronchen:2015vfa, Ronchen:2014cna, Kamano:2013iva,
JuBo, BnGa, Anisovich:2013tij, Kent, Doring:2010ap, Tiator:2010rp, EBAC, Vrana,
Giessen, Shklyar:2005xg} have usually fitted $\pi N$ amplitudes, derived from
previous  studies~\cite{WI08,
fit06,fit95,coulomb,hoehler,cutkosky1,cutkosky2,cutkosky3},  together with
reaction data. The fitted amplitude pseudo-data have either been taken from
single-energy analyses (SE) or energy-dependent (ED) fits covering the resonance
region. The SE analysis amplitudes, derived from fits to narrow energy bins of
data, have associated errors which have been used in the multi-channel fits, or
enlarged when these fits have become problematic. The smoother ED amplitudes
have also been taken at discrete energies, typically with subjective errors not
derived from the fit to data.   

There are several problems associated with fits to amplitude pseudo-data, which
we have attempted to address in this work. The most obvious of these is the fact
that the goodness of fit to these sets of amplitudes cannot be translated into a
quality of fit to the underlying dataset.  The subsequent comparison to
experimental $\pi N$ data may result~\cite{chinese}  in poorer than expected
agreement. In addition, uncertainties on the SE  amplitudes~\cite{WI08,
fit06,fit95, coulomb} do not account for correlated errors, which can be
substantial in some cases. 

In baryon spectroscopy, based on multi-reaction analysis, this has unwanted side
effects.  First, a statistical analysis of fit results is difficult if one of
the input channels is  not given by data. Second, as a consequence, the
significance of resonance signals, detected  in such multi-reaction fits, is
difficult to quantify. Consider, for example, the situation  in which an
additional resonance term leads to considerable improvement in the description 
of kaon photoproduction data. The description in the $\pi N\to\pi N$ reaction
might then  barely change. Indeed, one of the main motivations for the baryon
photoproduction program  is to search for missing states with small $\pi N$
resonance couplings. Yet, there will be a  non-zero impact in the description of
the $\pi N\to\pi N$ reaction. As long as that small  change in $\chi^2$ cannot
be tested in terms of statistical criteria, based on $\pi N$ data,  the
significance of the proposed new state will be difficult to assess.

In a similar way, Chiral Perturbation Theory (CHPT) and its unitary extension
(UCHPT) may profit  from an improved representation of SE amplitudes.  The
relevance of elastic $\pi N$ scattering partial waves for chiral dynamics, to
study the $\pi N$ $\sigma$-term, isospin breaking, or to obtain a quantitative
measure of low-energy constant (LEC) uncertainties, is reflected in the
literature~\cite{Bernard:1992qa, Fettes:1998ud, Meissner:1999vr, Fettes:2000xg, Fettes:2000bb,
Becher:2001hv, Fuchs:2003qc, Hoferichter:2009gn, Gasparyan:2010xz,
Alarcon:2011zs, Alarcon:2012kn, Chen:2012nx, Hoferichter:2015dsa, Hoferichter:2015tha,
Hoferichter:2015hva}.  Recently, several groups have begun to fit
low-energy $\pi N$ data directly~\cite{Wendt:2014lja,siemens}.

In UCHPT, the focus lies less on spectroscopy than the understanding of
resonance dynamics and its  nature in terms of hadronic components. Usually,
the $S$-wave amplitudes $S_{11}$ and $S_{31}$ are  subjects of interest. For
example, in Ref.~\cite{Bruns:2010sv} the $S_{11}$ and $S_{31}$ partial waves
were  fitted up to the energy of the $N(1535)S11$ resonance and the
$N(1650)S11$ emerged.  Furthermore, with the same hadronic amplitude, pion- and
$\eta$ photoproduction could be predicted~\cite{Mai:2012wy, Ruic:2011wf}.  The
role of chiral dynamics in $S$-wave baryonic resonances, including fits to $\pi
N$ partial waves, has been studied by many groups~\cite{Kaiser:1995cy,
Nieves:2001wt, Inoue:2001ip, Jido:2007sm, Doring:2008sv, Doring:2009qr,
Doring:2009uc, Doring:2013glu,  Khemchandani:2013nma, Garzon:2014ida}.  Other
examples, in which fits to $\pi N$ partial waves are crucial to investigate
chiral dynamics and to test models,  include the $D_{33}$ partial
wave~\cite{Kolomeitsev:2003kt, Sarkar:2004jh, Doring:2005bx, Doring:2007rz} and
a family of $J^P=1/2^-,\,3/2^-$ states~\cite{Oset:2009vf, Khemchandani:2013nma,
Garzon:2014ida}.  Clearly, an improved representation of $\pi N$ data beyond SE
amplitudes will lead to a more reliable  determination of LECs, and thus, more
reliable predictions of other hadronic reactions within UCHPT. 

In summary, SE $\pi N$ amplitudes represent the test ground for a wide range of
theory and models from baryon spectroscopy and chiral resonance dynamics to
tests of quark models~\cite{Kiswandhi:2003ca, An:2011sb, Golli:2011jk}. 
Attaching more statistical meaning to those solutions would considerably
advance the understanding  of hadron dynamics.

The aim of this paper is to provide an easy-to-implement representation of the
$\pi N\to\pi N$ data in terms of covariance matrices and best $\chi^2$ values
for each set of SE amplitudes. With this,  the $\pi N\to\pi N$ reaction can be
included in multi-reaction spectroscopy fits in a statistically more meaningful
way through correlated $\chi^2$ fits. The effect of systematic errors
associated with the underlying data provides a subtle difficulty which we
discuss in detail below. 

Together with this manuscript, numerical values for matrices and $\chi^2$
values are provided on the SAID~\cite{SAID} and JPAC~\cite{JPAC} web pages for further use.


\section{Generating SE amplitudes}

In the following, we restrict our attention to the single-energy (SE)
amplitudes, which are generated starting from a global, energy-dependent (ED)
fit, and give a better fit to data. These amplitudes show more scatter than
would appear in the ED fit. This is preferable in a multi-channel analysis which
may interpret apparently random fluctuations in the single-channel fit as
resonance signatures. Here, we use the most recent ED fit of Ref.~\cite{WI08}. 

Data for each of the SE analyses have been taken from the SAID
database~\cite{SAID} with an energy interval depending on the density of
experimental measurements. This interval varies from 2 MeV, for the low-energy
region, to 50 MeV, at the highest energies where data are sparse.  A finite binning
in energy increases the number of data constraints but requires an assumption for
the energy dependence, which is taken to be linear. The quoted amplitudes
correspond to the central energy. The $\chi^2$ fit to data is carried out, using
the form
\begin{equation}
\chi^2 = \sum_i \left( {{N \Theta_i - \Theta_i^{\rm exp} }\over {\epsilon_i}}
\right)^2 + \left( {{N-1}\over{\epsilon_N}} \right)^2
\label{chi0}
\end{equation}  
where $\Theta_i^{\rm exp}$ is an experimental point in an angular distribution
and $\Theta_i$ is the fit value. Here the overall systematic error,
$\epsilon_N$, is used to weight an additional $\chi^2$ penalty term due to 
renormalizaton of the fit by the factor $N$. The statistical error is given by
$\epsilon_i$.  It has been shown that the above renormalization factors can be
determined at each  search step and do not have to be explicitly included in the
search~\cite{Arndt66}. Empirical renormalization factors have also been used in
fits to low-energy data based on chiral perturbation theory~\cite{siemens}.

The search is stabilized in two ways. Clearly, one cannot search an infinite
number of partial waves. As a result, the number of included waves is determined
by their contribution to the cross section, with all higher waves being taken
from the ED fit. In addition, ED amplitude pseudo-data are included in the fit,
with large uncertainties,  to keep the SE solution in the neighborhood of the ED
result. Clearly, with overly tight constraints, one could generate a SE fit
arbitrarily close to the ED value. However, in practice, the constraints allow
sufficient freedom and contribute very little (less than 1\%) to the total
$\chi^2$. The searched waves are elastic until their
contribution to the reaction cross section is significant, as determined in  the
ED analysis.


\section{Using the error matrix}
A pion-nucleon partial wave $f_i$ is parametrized by two real parameters. Here,
we choose the phase shift $\delta_i$ and $\rho_i$ where 
\be
\cos\rho_i= \eta_i\ ,
\label{rho}
\ee
with elasticity parameter $\eta_i$ and the scattering amplitude 
\be
{\rm Re}\, f_i&=&\frac{1}{2}\cos\rho_i\sin(2\delta_i),\non
{\rm Im}\, f_i&=&\frac{1}{2}\left(1-\cos\rho_i\cos (2\delta_i)\right) \ .
\label{f}
\ee
In the following, the set of parameters for a given set of partial waves is
called generically $A_i$, ordered in a vector ${\bf A}$. The $\chi^2$ of a SE
solution can be expanded around the minimum at ${\bf A}=\hat {\bf A}$,
\be
\chi^2({\bf A})&=&\chi^2(\hat {\bf A})
+({\bf A}-\hat{\bf A})^T \hat\Sigma^{-1} ({\bf A}-\hat{\bf A})\nonumber \\
&+&{\cal O}({\bf A}-\hat{\bf A})^3 ,
\label{chi1}
\ee
where $\hat {\bf A}$ is the estimate of the partial waves from data and
$\hat\Sigma$ is the estimate of the covariance matrix.  A {\it correlated
$\chi^2$ fit} to a SE solution means the use of the same Eq.~(\ref{chi1}) for
the $\chi^2$ up to ${\cal O}({\bf A}^2)$, in particular of the full covariance
matrix and not only its diagonal elements given by the partial-wave variances
$(\Delta A_i)^2$. Thus, using $\hat\Sigma$ and $\chi^2(\hat {\bf A})$ of this
paper in a correlated $\chi^2$ fit provides in principle the same $\chi^2$ as
fitting to the actual data up to ${\cal O}({\bf A}^2)$, resolving the issues
raised in the Introduction.

In an actual correlated $\chi^2$ fit, either $(\delta_i,\rho_i)$ may be fitted,
using the quoted covariance matrices, or, the possibly more familiar scattering
amplitudes $({\rm Re}\,f_i,{\rm Im}\,f_i)$ may be utilized, requiring a
transformation of the covariance matrices,
\be
\hat\Sigma_f=Q^T\hat\Sigma Q,
\ee
where $Q$ is a block-diagonal matrix $Q={\rm diag}(Q_j)$ with
\be
Q_j=
\left(
\begin{array}{cc}
\cos\rho_i\cos(2\delta_i) & \cos\rho_i\sin(2\delta_i)\\
-\frac{1}{2}\sin\rho_i\sin(2\delta_i)&\frac{1}{2}\sin\rho_i\cos(2\delta_i)
\end{array}
\right) \ ,
\ee
for inelastic partial waves, with $\rho_i\neq 0$, and 
\be
Q_j=
\left(
\begin{array}{cc}
\cos(2\delta_i)& \sin(2\delta)
\end{array}
\right) \ ,
\ee
for the elastic partial waves (note that $Q$ is not necessarily a square
matrix). For groups accustomed to fitting the amplitudes $f_i$, it may be more
convenient in practice to evaluate $(\delta_i,\rho_i)$ using Eq.~(\ref{rho}) and
inverting Eqs.~(\ref{f}) to fit to the quoted covariance matrices directly. 


\subsection{Format of covariance matrices}

The format of covariance matrices $\hat\Sigma$ and $\chi^2$ estimates
$\chi^2(\hat {\bf A})$ are specified on the SAID web page~\cite{SAID}. At the
time of publication,  we quote the parameters corresponding to the WI08
solution~\cite{WI08}. The web page will be updated as new data are produced
and analyzed.
Along with the necessary parameters to carry out correlated $\chi^2$ fits,
simple subroutines are provided to read the
parameters into suitable variables. The parameters to describe the $\chi^2$ are:
central $W$ of the energy bin of a given SE solution, ordering of partial wave
$\delta_i$ and $\rho_i$ parameters according to isospin $I$, orbital angular
momentum $L$, total angular momentum $J$, and the actual values of $\hat {\bf
A}$, $\chi^2(\hat {\bf A})$ and $\hat\Sigma$, in the given ordering.
Additionally, the number of data points in the bin is quoted.


\subsection{Representation of the $\chi^2$}
As discussed, $\hat {\bf A}$, $\chi^2(\hat {\bf A})$ and $\hat\Sigma$ for SE
solutions provide the necessary input for other groups to carry out fits with a 
$\chi^2$ that represents, in principle, the $\chi^2$ of a direct fit to $\pi N$
data. A few remarks concerning the advantages and limitations of this method are
in order.
\begin{itemize}[leftmargin=*, nosep]
\item
Non-linear contributions. As discussed following Eq.~(\ref{chi1}), a correlated
$\chi^2$ fit captures only the quadratic terms in the expansion around the
minimum. Non-linear corrections of ${\cal O}({\bf A}^3)$ are neglected. Testing
selected covariance matrices,  we found that non-linear corrections only become
relevant far beyond the parameter region over  which a fit is considered to be
good. In Sec.~\ref{sec:example} an explicit example is discussed.
\item
Finite bin width. As mentioned, the bin widths become up to 50~MeV wide at the
largest energies.  However, partial-wave solutions have a smooth energy
dependence, and single-energy solutions are  allowed to vary linearly within a
bin. The impact on the $\chi^2$ from finite bin width is not significant and
only central values of the bins are quoted.
\item
Electromagnetic corrections. As the SE solutions are corrected using the method
described in detail in Ref.~\cite{coulomb},  other groups using the present
results do not have to implement electromagnetic corrections required to fit the
data. Conversely, the implementation of electromagnetic corrections cannot be
altered without a re-fit to the data.

\item
Renormalization. The SE solutions are obtained by allowing for a multiplicative
renormalization  according to Eq.~(\ref{chi0}). Any group using the present
results implicitly accepts the normalization  obtained in the SAID analysis of
elastic $\pi N$ scattering. Beyond this, no additional renormalization  can be
performed in correlated $\chi^2$ fits. The effect of renormalization becomes
increasingly  relevant when moving away from the estimated $\chi^2$ minimum at
${\bf A}=\hat{\bf A}$.  We discuss a typical example in Sec.~\ref{sec:example}.

The effect from renormalizations ``frozen'' at the SAID SE solution value at
${\bf A}=\hat{\bf A}$  represents the largest difference between the correlated
and the actual $\chi^2$, in which renormalization  is dynamically adapted for
any ${\bf A}$. Yet, as renormalization tends to be small to moderate, and for 
${\bf A}$  in the vicinity of $\hat{\bf A}$, the effect can be neglected.
\end{itemize}
In summary, there are advantages in using the present fit method over a direct
fit to data  (no need to implement electromagnetic corrections), but also
limitations. Especially if a correlated  $\chi^2$ fit is poor, i.e., with
parameters ${\bf A}$ far away from $\hat{\bf A}$, the correlated and  actual
$\chi^2$ can be quite different. In that case, one can only resort to a direct
fit to data,  allowing for dynamic renormalization. Then, the fit function must
be renormalized, rather than the data,  to avoid the bias discussed in
Ref.~\cite{D'Agostini:1993uj}. See also Ref.~\cite{Ball:2009qv} for a further
discussion of the topic. 

With the limitations discussed, correlated $\chi^2$ fits still represent a much
improved treatment of  the elastic $\pi N$ reaction, compared to uncorrelated
fits to SE solutions, as available up to now.  This will be demonstrated in an
example in the next section.
  

\section{An explicit example}
\label{sec:example}

\begin{table}
\begin{center}
\begin{tabular}{|c|c|c|c|}
\hline
 90 MeV SE	& WI08 (ED)	& WI08 (SE)	& WI08 (SE-No Renorm)	\\
 (87-92) MeV	& 		&		&			\\
\hline
S$_{11}$	& 8.43 		& 8.11(0.11) 	& 8.02(0.11)		\\
S$_{31}$	&  -8.21 	& -8.11(0.10) 	& -7.68(0.10)   	\\
P$_{11}$      	& -1.01  	& -0.71(0.09)   & -0.58(0.09)        	\\
P$_{33}$      	& 17.31   	& 17.16(0.05)  	& 16.68(0.05)     	\\
\hline
$\chi^2$/data	& 150/121    	& 124/121   	& 301/121		\\
\hline
\end{tabular}
\caption{\label{tab:chi} 
Fits to data near $T_\pi=90$~MeV. Quoted are the phase shifts $\hat \delta_i
$(deg).  WI08~\cite{WI08} is the energy-dependent (ED) fit, SE is the
single-energy fit, allowing renormalization, based on the ED fit. The last
column gives a SE fit without allowing renormalization of the fit (see text).
}
\end{center}
\end{table}

Table \ref{tab:chi} compares fits to data with lab pion kinetic energies $T_\pi$
between 87 and 92 MeV. Quoted are the phase shifts $\hat A_i=\hat
\delta_i $(deg). The fit WI08~\cite{WI08} is an ED parametrization  of data
covering the full resonance region (second column). It employs a normalization
of the fit function. 
Smaller partial waves, present in the ED solution but not searched in the SE
fit, are omitted from the table. 

From this starting point, the most important partial waves have been searched to
fit data in the chosen energy bin. In this case, 
$S_{11},\,S_{31},\,P_{11}$, and $P_{33}$ phase shifts have been searched with
other parameters  held fixed at WI08 values.  This is the SE fit in the third
column quoted with errors determined from the corresponding diagonal elements of
the covariance matrix. As a simpler point of comparison, a second SE fit has
been done without allowing for renormalization of the fit
(last column). Here the fit is significantly worse.

\begin{figure}
\begin{center}
\includegraphics[width=0.92\linewidth]{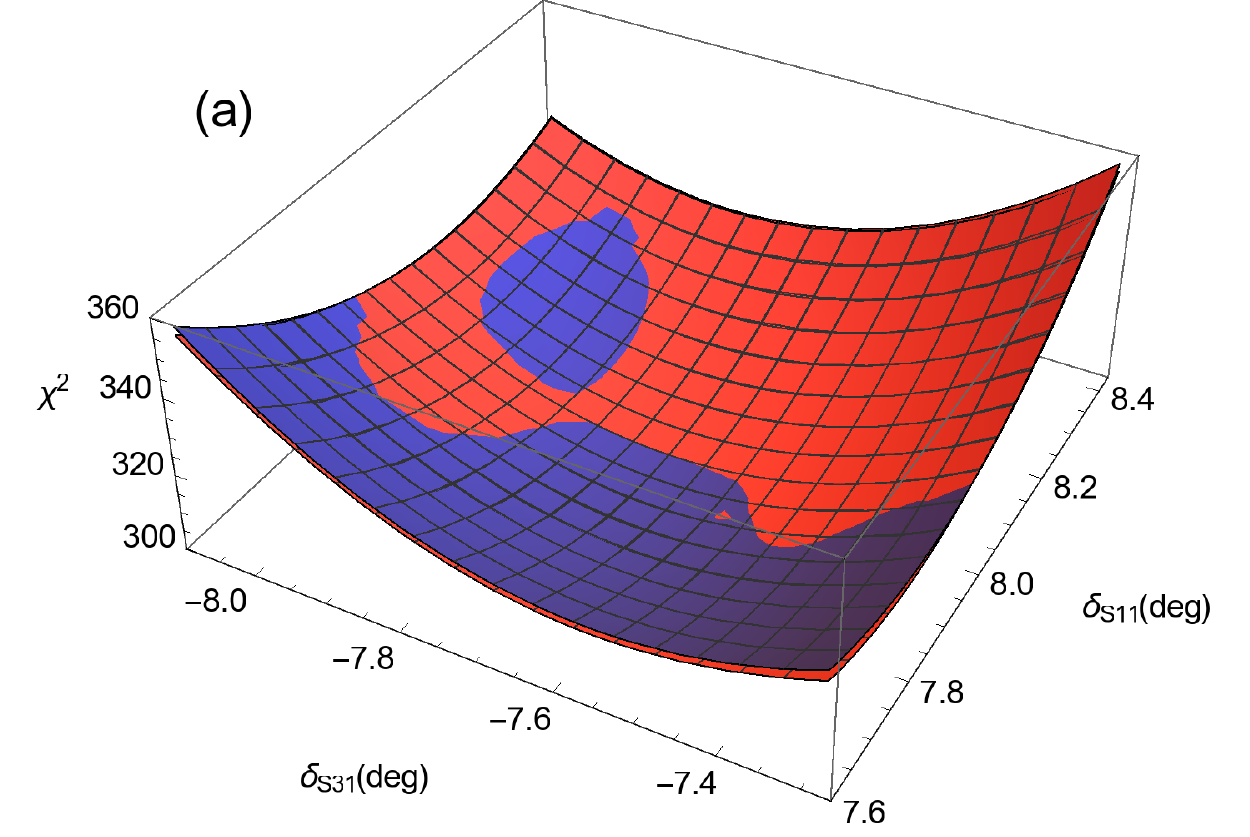}\\
\includegraphics[width=0.92\linewidth]{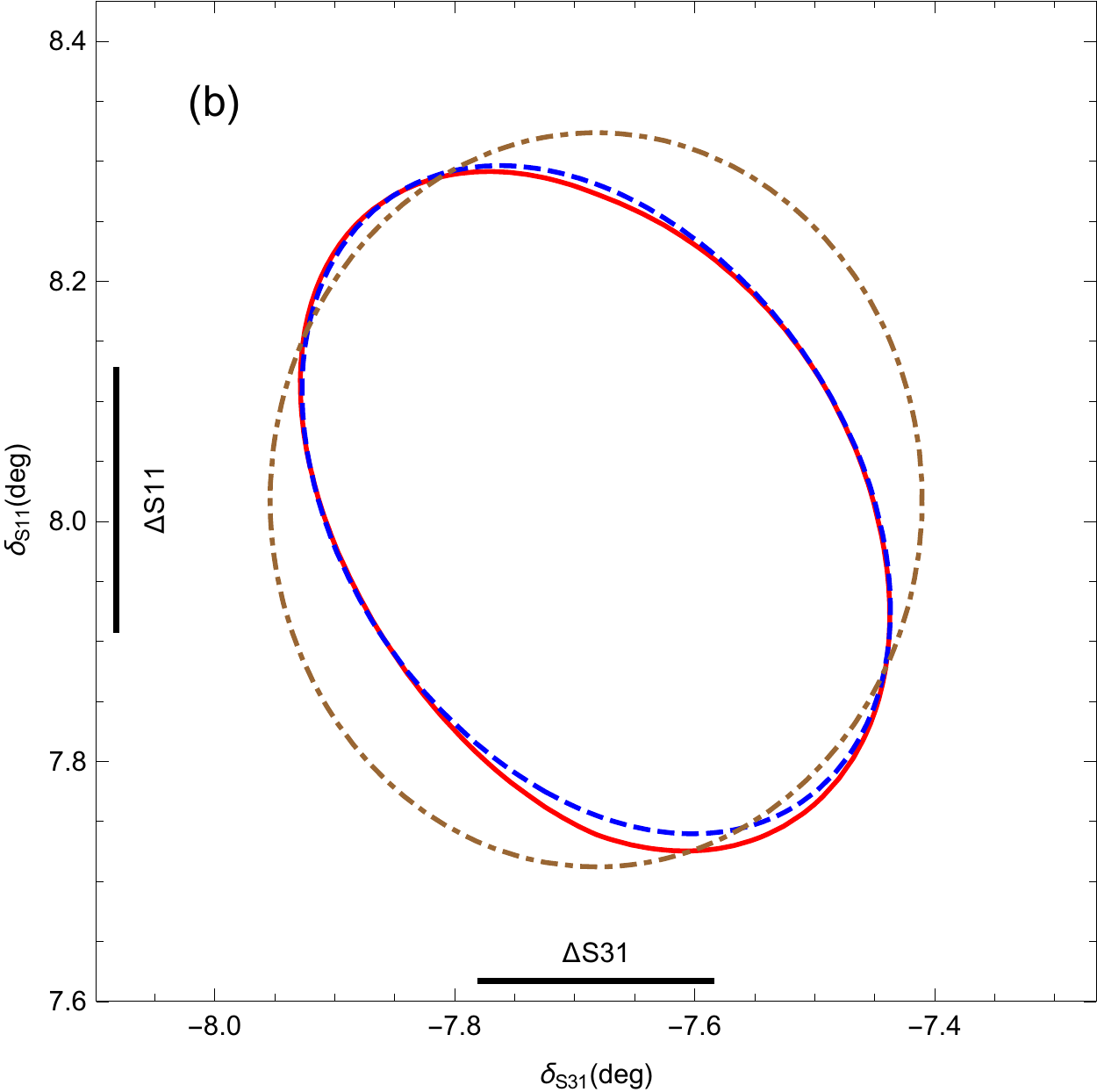}
\end{center}
\caption{
The $\chi^2$ without renormalization (last column of Table~\ref{tab:chi}). 
(a) The $\chi^2$ of the SES
for $T_\pi\in[87,92]$~MeV as a function of $\delta_{S11}$ and $\delta_{S31}$
with the values of all other partial waves fixed at the minimum. The red (blue)
surface shows the actual $\chi^2$ (the $\chi^2$ predicted from the covariance
matrix). (b) Contours of constant $\Delta\chi^2=8$ for the
actual $\chi^2$ (solid red), the $\chi^2$ predicted from the full covariance
matrix (dashed blue), and from the covariance matrix neglecting correlations
(dash-dotted brown line). Parameter errors $\Delta S11$, $\Delta S31$ are
indicated with bars.
}
\label{fig:chi1}
\end{figure}
Starting from this last SE fit, and its best $\chi^2$, we see from
Eq.~(\ref{chi1}) that the $\chi^2$ should increase quadratically as one moves
away from the minimum. In the top panel of Fig.~\ref{fig:chi1}, we compare the
$\chi^2$ variation for the two $S$-wave amplitudes as given by the corresponding
error matrix and an actual fit to data  (the other two partial waves are held at
their best values $\hat \delta_{P11}$ and $\hat \delta_{P33}$).  Shown is a
region well beyond the $\Delta\chi^2=2.30$ ellipse that marks the 68\%
confidence region of  a two-parameter fit (and well beyond the
$\Delta\chi^2=4.72$ ellipse of a 4-parameter fit).  The parabolic behavior of
the correlated $\chi^2$ predicts well the actual $\chi^2$ within the shown
region.  Thus, the ${\cal O}({\bf A}^3)$ corrections of Eq.~(\ref{chi1}) are
indeed very small well beyond the  region in which a fit can be considered good.
 
In the bottom panel of Fig.~\ref{fig:chi1} we show the
$\Delta\chi^2(\hat\Sigma)=8$ ellipse from $\hat\Sigma$ (solid, red) and compare
with the actual $\Delta\chi^2=8$ line (dashed, blue). The figure shows again
that the covariance matrix predicts the rise of the $\chi^2$ well. For example,
at $(\delta_{S11}, \delta_{S31})=(8.42\,\text{deg},-7.28\,\text{deg})$ the
difference between $\Delta\chi^2(\hat\Sigma)$ and the actual $\Delta\chi^2$ is
only 2, compared to an absolute scale given by $\chi^2=359$ at this point. Along
the axes, the figure also shows the parameter errors, given by the maximal
extensions of the $\Delta\chi^2=1$ ellipse. 

In addition, a $\Delta\chi^2=8$ error ellipse is shown that is obtained from the
covariance $\hat\Sigma_0$ in which all off-diagonal elements are set to zero,
i.e., ignoring correlations (dash-dotted, brown). The effect is sizeable: At
$(\delta_{S11}, \delta_{S31})$ considered before one has
$\Delta\chi^2(\hat\Sigma)=56$ and $\Delta\chi^2(\hat\Sigma_0)=31$, i.e., only
$55\%$ of the correlated value. At higher energies, where parameters are
generally more strongly correlated, this discrepancy becomes much larger. 

The breakdown of $\chi^2$ contributions is then as follows: the $\chi^2$
at the minimum is $\chi^2(\hat{\bf A})$=301, the contribution from correlations
amounts to $\Delta\chi^2=56$, and the sum $\chi^2=357$ is 0.5\% different from
the actual $\chi^2$ found from a comparison to data. 
In contrast, if one had mistakenly regarded the SE
solutions as uncorrelated data points (as done in some analyses), a meaningless
$\chi^2$=31 would have been obtained at $(\delta_{S11},
\delta_{S31})=(8.42\,\text{deg},-7.28\,\text{deg})$.

To conclude this section, the effects of normalization are discussed.
Recall that the minimum at ${\bf A}=\hat{\bf A}$ in the standard SE fit (third
column of Table~\ref{tab:chi}) is obtained allowing for renormalization of the
minimizing function. The covariance matrix is then numerically estimated from
the Hessian, $\hat\Sigma=2H^{-1}$ with $H_{ij}=\partial^2\chi^2/(\partial
A_i\partial A_j)$, using the penalized $\chi^2$ from Eq.~(\ref{chi0}), i.e.,
including the renormalization. To that end, the
covariance matrix includes information about the change in normalization when
moving away from the minimum, but with a value ``frozen'' at the minimum.
Moving away from the minimum, both the fitted amplitudes
and the fit function normalization factors work to reduce the
$\chi^2$, resulting in a non-quadratic variation. However, if one is close to
the minimum, the error matrix should still give a reasonable estimate of the
data $\chi^2$. 

\begin{figure}
\begin{center}
\includegraphics[width=0.92\linewidth]{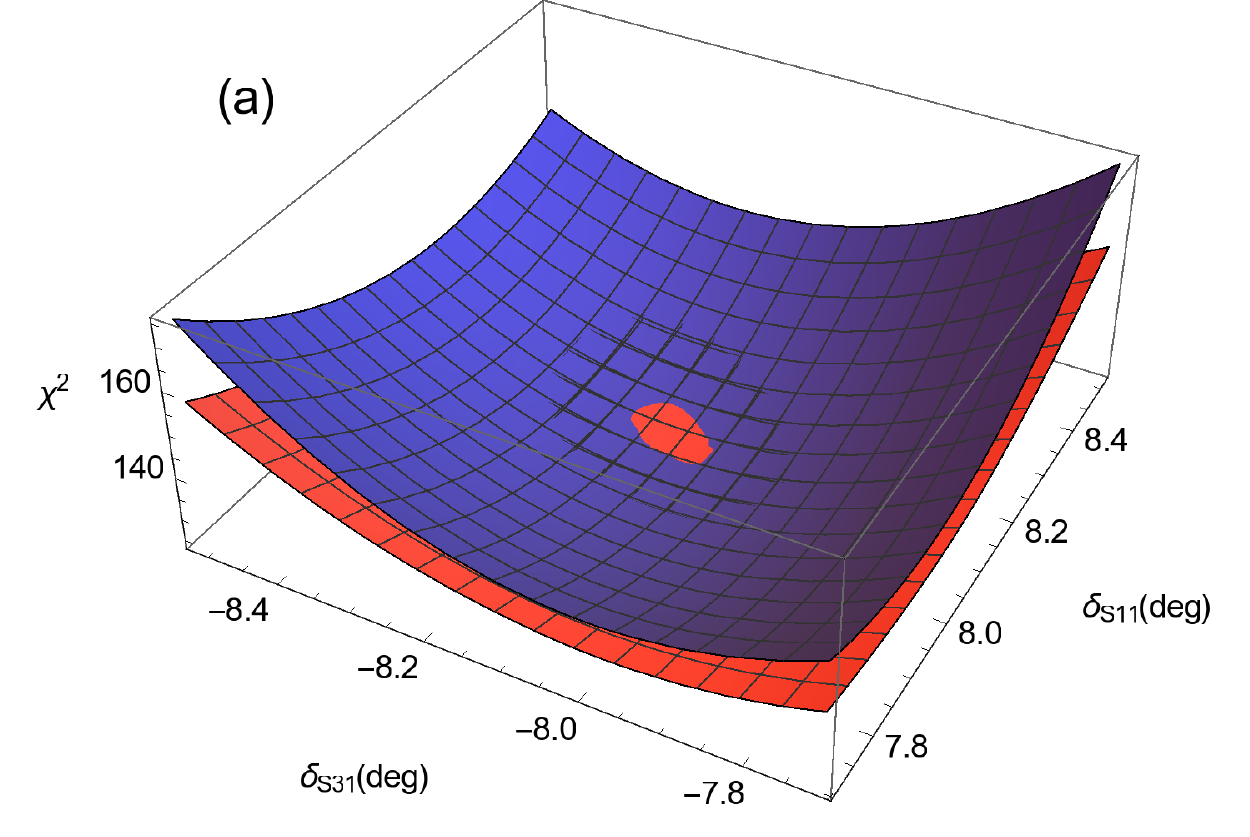}\\
\includegraphics[width=0.92\linewidth]{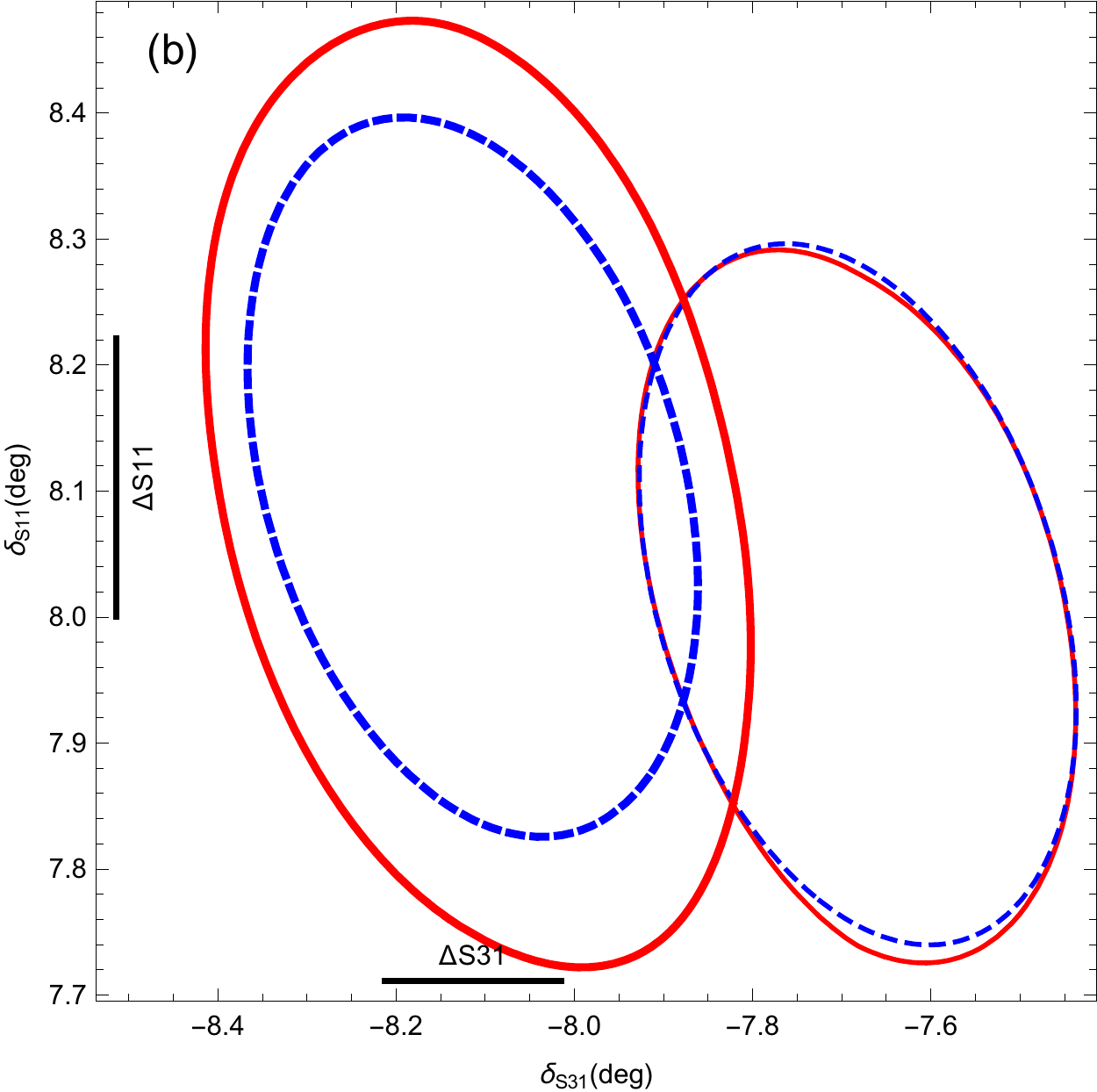}
\end{center}
\caption{
The $\chi^2$ with renormalization. Notation as in Fig.~\ref{fig:chi1}.
(a) The red (blue) surface shows the actual $\chi^2$ with
renormalization (the $\chi^2$ predicted from the covariance matrix).
(b) Contours of constant $\Delta\chi^2=8$ for the actual
$\chi^2$ with renormalization (thick solid red), the $\chi^2$ predicted from the
covariance matrix (thick dashed blue), and the case without renormalization from
Fig.~\ref{fig:chi1} (thin lines).
}
\label{fig:chi2}
\end{figure}
In Fig.~\ref{fig:chi2}, bottom panel, $\Delta\chi^2$ curves from the
normalizable SE solution (thick) lines are shown. The curves from the previously
discussed case (no normalization) are re-plotted for comparison (thin lines).
The thick solid red (thick dashed blue) lines show the actual
$\Delta\chi^2$ values (the $\Delta\chi^2$ values predicted from the covariance
matrix). We observe larger deviations of the actual $\chi^2$ from the predicted
one, that are a consequence of the discussed dynamic normalization, changing at
any point in parameter space for the evaluation of the actual $\chi^2$. Note,
however, that this example has been chosen for the $\Delta \chi^2=8$ contour,
i.e., far away from the minimum. There, a maximal deviation of actual and
predicted $\chi^2$ of 5 \% is observed.

\begin{figure}
\begin{center}
\includegraphics[width=0.75\linewidth]{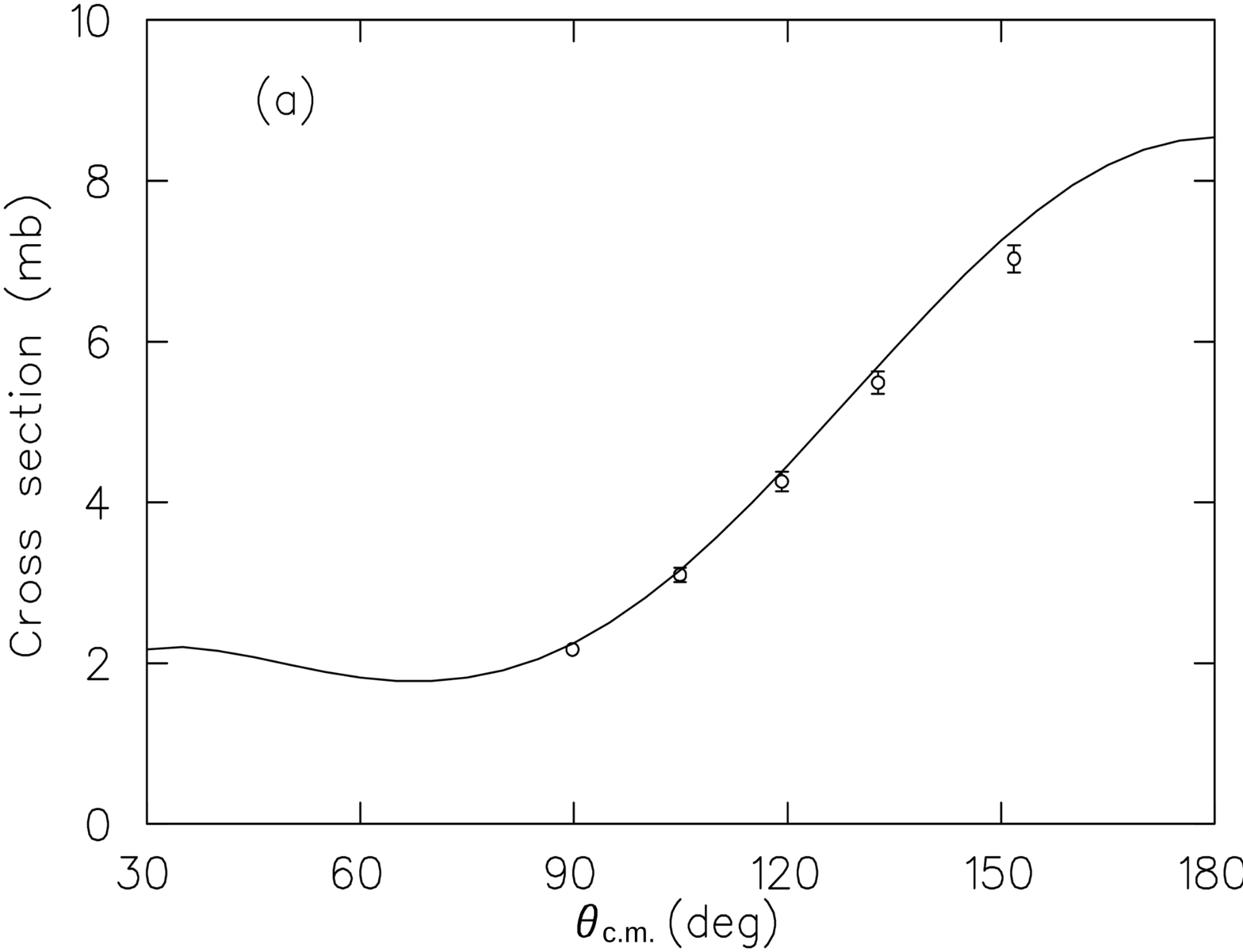}\\
\includegraphics[width=0.75\linewidth]{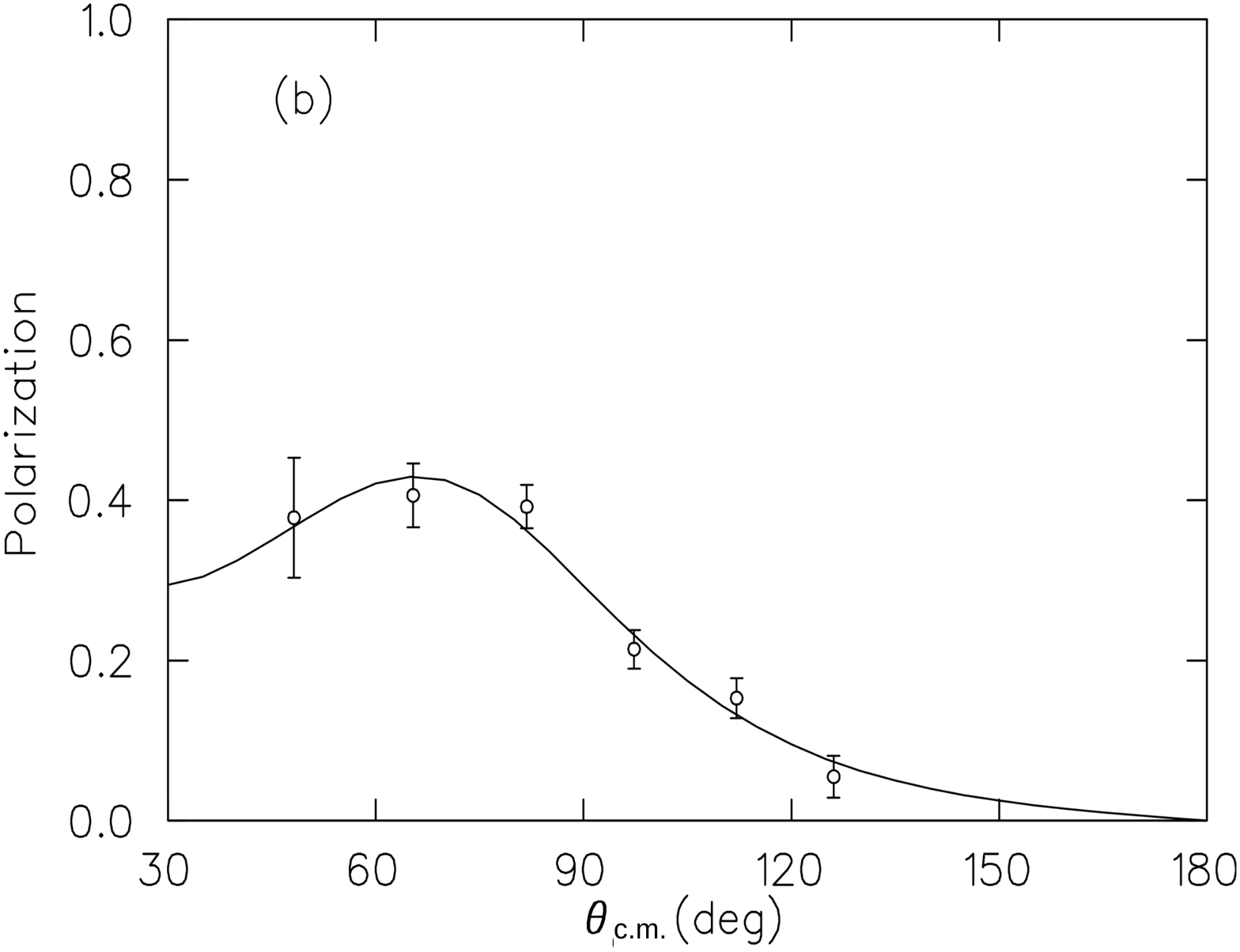}
\end{center}
\caption{(a) Differential cross section at $T_\pi=91.7$~MeV and $\pi^+p$ data of
Ref.~\cite{dsg_data}.  (b) Polarization (P) at $T_\pi=87.2$~MeV and $\pi^+p$
data of Ref.~\cite{p_data}.  The 90 MeV SE fit is shown; the normalization
$N$ from Eq.~(\ref{chi0}) acquires a value of $N=0.98$ for the differential
cross section (not applied in figure).
}
\label{fig:data}
\end{figure}
For further illustration, Fig.~\ref{fig:data} shows a selection of data from the
considered $T_\pi=87-92$~MeV energy bin and the SE fit obtained allowing
normalization. The effect of normalization is visible for the differential cross
section, which acquires a normalization factor of 0.98, constrained by the
penalty term in Eq.~(\ref{chi0}). The factor, not applied in the figure, shifts
the curve closer to the data, significantly reducing the $\chi^2$.


\subsection{Fits with fewer parameters}
Some theory or model approaches describe fewer partial waves than provided
in the covariance matrices. For example, chiral unitary approaches are often
restricted to the lowest partial waves. How should one use the
covariance matrices in these cases? 

As an example, assume that model M describes $\delta_{S11}$, while the
covariance matrix comprises $\delta_{S11}$ and $\delta_{S31}$ (see, e.g., the
figures of this section). Suppose $\delta_{S11}$ in model M takes the value
$\delta_{S11}=\hat \delta_{S11}+\Delta S11$. In the $\delta_{S11}, \delta_{S31}$
space, this corresponds to the right vertical tangent to the $\Delta\chi^2=1$
ellipse. Then, there exists one value $\delta_{S31}$ such that indeed
$\Delta\chi^2=1$. On the other hand, marginalizing the bivariate distribution
over $\delta_{S31}$, one obtains a Normal distribution with variance $(\Delta
S11)^2$, corresponding to a covariance matrix $\hat\Sigma=(\Delta S11)^2$.
According to that reduced covariance matrix, the $\Delta\chi^2$ at
$\delta_{S11}=\hat \delta_{S11}+\Delta S11$ has also increased by one,
$\Delta\chi^2=1$. In summary, fitting the reduced covariance matrix is
equivalent to fitting the entire covariance matrix, with $\delta_{S11}$ coming
from model M, and optimizing all other parameters simultaneously. (Within M one
cannot make any statement about the size of these other parameters/partial
waves.)

The generalization to several parameters is straightforward. It can be shown
that the reduced covariance matrix after marginalization is given by simply
eliminating, from the full covariance matrix, the rows and columns corresponding
to the marginalized parameters. Then, model M with fewer partial waves is fitted
to that reduced matrix, and the unchanged $\chi^2(\hat {\bf A})$ is added
according to Eq.~(\ref{chi1}).


\section{Summary and Conclusions}
Covariance matrices and other fit properties of the SAID SE solutions are
provided to allow other groups to carry out correlated $\chi^2$ fits to the
elastic $\pi N$ scattering reaction. In principle, the obtained $\chi^2$ is then
a good approximation to the $\chi^2$ one would obtain if fitting 
directly to experimental data.
This has various practical advantages: Coulomb corrections are not an issue and
normalization factors are included. However, the latter bear some subtleties as
discussed. Furthermore, when fitting to SAID SE solutions, in the proposed
manner, one implicitly accepts the chosen bin width and omission of non-linear
contributions to the $\chi^2$ beyond the covariance matrix. In practice, we
found these effects to be negligible, with the largest discrepancies coming from
normalization. However, it has been checked that, close to the minimum, this 
effect is under control.

With correlated $\chi^2$ fits, it is now possible to  fit the SAID SE solutions
in a statistically meaningful way.  For baryon spectroscopy, this is a
prerequisite to quantify the significance of resonance signals, usually
performed in multi-reaction fits in which, so far, the precise statistical
impact of $\pi N$ partial waves has been unknown. Other approaches, such as
quark-model calculations, CHPT, or unitary extensions thereof can also benefit
from the proposed fitting scheme, allowing, e.g., for an improved determination
of low-energy constants. The numerical input needed to carry out correlated
$\chi^2$ fits is provided online.


\begin{acknowledgments}

This work was supported in part by the U.S. Department of Energy Grant
DE-SC0014133, by the National Science Foundation (CAREER grant No. 1452055, PIF
grant No. 1415459), by GWU (startup grant), and by the DFG and NSFC through the
Sino-German CRC 110. 
M.D. is also supported by the U.S. Department of Energy, Office of Science, 
Office of Nuclear Physics under contract DE-AC05-06OR23177. 
The authors gratefully acknowledge the computing time
granted on the supercomputer JURECA at J\"ulich Supercomputing Centre (JSC) and
thank C\'esar Fern\'andez Ram\'irez for useful discussions.

\end{acknowledgments}


\end{document}